\documentclass[runningheads]{llncs}

\usepackage{graphicx}
\usepackage{floatrow}
\usepackage{fancyhdr}

\begin{document}

\title{QURATOR: Innovative Technologies for Content and Data Curation}

\author{
Georg Rehm\inst{1}\and 
Peter Bourgonje\inst{1}\and 
Stefanie Hegele\inst{1}\and 
Florian Kintzel\inst{1}\and \\
Juli\'{a}n Moreno Schneider\inst{1}\and 
Malte Ostendorff\inst{1}\and 
Karolina Zaczynska\inst{1}\and \\
Armin Berger\inst{2}\and 
Stefan Grill\inst{2}\and
S\"oren R\"auchle\inst{2}\and 
Jens Rauenbusch\inst{2}\and \\
Lisa Rutenburg\inst{2}\and 
Andre Schmidt\inst{2}\and 
Mikka Wild\inst{2}\and 
Henry Hoffmann\inst{3}\and \\
Julian Fink\inst{3}\and
Sarah Schulz\inst{3}\and 
Jurica \v{S}eva\inst{3}\and
Joachim Quantz\inst{4}\and \\
Joachim B\"{o}ttger\inst{4}\and 
Josefine Matthey\inst{4}\and 
Rolf Fricke\inst{5}\and 
Jan Thomsen\inst{5}\and \\
Adrian Paschke \inst{6}\and 
Jamal Al Qundus\inst{6}\and 
Thomas Hoppe \inst{6}\and 
Naouel Karam \inst{6}\and \\
Frauke Weichhardt\inst{7}\and 
Christian Fillies\inst{7}\and 
Clemens Neudecker\inst{8}\and
Mike Gerber \inst{8}\and
Kai Labusch\inst{8}\and
Vahid Rezanezhad \inst{8}\and
Robin Schaefer \inst{8}\and
David Zellh\"{o}fer \inst{8}\and \\
Daniel Siewert\inst{9}\and 
Patrick Bunk\inst{9}\and 
Julia Katharina Schlichting\inst{9}\and \\
Lydia Pintscher\inst{10}\and 
Elena Aleynikova\inst{10}\and
Franziska Heine\inst{10}}

\authorrunning{Georg Rehm et al.}

\institute{
DFKI GmbH, Alt-Moabit 91c, 10559 Berlin, Germany
\and 
3pc GmbH Neue Kommunikation, Prinzessinnenstra\ss{}e 1, 10969 Berlin, Germany
\and
Ada Health GmbH, Adalbertstra\ss{}e 20, 10997 Berlin, Germany
\and
ART+COM AG, Kleiststra\ss{}e 23-26, 10787 Berlin, Germany
\and 
Condat AG, Alt-Moabit 91d, 10559 Berlin, Germany
\and
Fraunhofer FOKUS, Kaiserin-Augusta-Allee 31, 10589 Berlin, Germany
\and
Semtation GmbH, Geschwister-Scholl-Stra\ss{}e 38, 14471 Potsdam, Germany
\and
Staatsbibliothek zu Berlin (SPK), Potsdamer Stra\ss{}e 33
10785 Berlin, Germany
\and
Ubermetrics Technologies GmbH, Kronenstra\ss{}e 1, 10117 Berlin, Germany 
\and
Wikimedia Deutschland e.\,V., Tempelhofer Ufer 23-24, 10963 Berlin, Germany
\\[1ex]
Corresponding author: Georg Rehm -- \email{georg.rehm@dfki.de}
}

\maketitle 

\begin{abstract}
In all domains and sectors, the demand for intelligent systems to support the processing and generation of digital content is rapidly increasing. The availability of vast amounts of content and the pressure to publish new content quickly and in rapid succession requires faster, more efficient and smarter processing and generation methods. With a consortium of ten partners from research and industry and a broad range of expertise in AI, Machine Learning and Language Technologies, the QURATOR project, funded by the German Federal Ministry of Education and Research, develops a sustainable and innovative technology platform that provides services to support knowledge workers in various industries to address the challenges they face when curating digital content. The project's vision and ambition is to establish an ecosystem for content curation technologies that significantly pushes the current state of the art and transforms its region, the metropolitan area Berlin-Brandenburg, into a global centre of excellence for curation technologies.

\keywords{Curation Technologies \and Language Technologies \and Semantic Technologies \and Knowledge Technologies \and Artificial Intelligence}
\end{abstract}

\thispagestyle{fancy}
\fancyhf{}
\fancyhead[R]{}
\fancyfoot[C]{\tiny Copyright \textcopyright\ 2020 for this paper by its authors.\\ Use permitted under Creative Commons License Attribution 4.0 International (CC BY 4.0).}
\renewcommand{\headrulewidth}{0pt}

\setcounter{footnote}{0}

\section{Introduction}

Digital content and online media have gained immense importance, especially in business, but also in politics and many other areas of society. Some of the many challenges include better support and smarter technologies for digital content curators who are exposed to an ever increasing stream of heterogeneous information they need to process further. For example, professionals in a digital agency create websites or mobile apps for customers who provide documents, data, pictures, videos etc.~that are processed and then deployed as new websites or mobile apps. Knowledge workers in libraries digitize archives, add metadata and publish them online. Journalists need to continuously stay up to date to be able to write a new article on a specific topic. Many more examples exist in various industries and media sectors (television, radio, blogs, journalism, etc.). These diverse work environments can benefit immensely from smart semantic technologies that help content curators, who are usually under great time pressure, to support their processes. Currently, they use a wide range of non-integrated, isolated, and fragmented tools such as search engines, Wikipedia, databases, content management systems, or enterprise wikis to perform their curation work. Largely manual tasks such as smart content search and production, summarization, classification as well as visualization are only partially supported by existing tools \cite{rehm2015c}.

The QURATOR project\footnote{\url{https://qurator.ai}}, funded by the German Federal Ministry of Education and Research (BMBF), with a project runtime of three years (11/2018-10/2021), is based in the metropolitan region Berlin/Brandenburg. The consortium of ten project partners from research and industry combines vast expertise in areas such as Language Technologies as well as Knowledge Technologies, Artificial Intelligence and Machine Learning. The project’s main goal is the development of a sustainable technology platform that supports knowledge workers in various industries. Non-efficient process chains increase the manual processing effort for workers even more. The platform will simplify the curation of digital content and accelerate it dramatically. AI techniques are integrated into curation technologies and curation workflows in the form of industry solutions covering the entire life cycle of content curation. The solutions being developed focus on curation services for the sectors of culture, media, health and industry.

In Section~\ref{sec:platform} we describe the emerging QURATOR technology platform. In the main part of this article, Section~\ref{sec:projects}, we provide brief summaries of the ten partner projects. Section~\ref{sec:summary} concludes the article with a short summary.

\section{The QURATOR Curation Technology Platform}
\label{sec:platform}

The centerpiece of the project is the development of a platform for digital curation technologies. The project develops, integrates and evaluates various services for importing, preprocessing, analyzing and generating content that covers a wide range of information sources and data formats, spanning use cases from several industries and domains. A special focus of the project is on the integration of AI methods to improve the quality, flexibility and coverage of the services. 

\begin{figure}
\includegraphics[width=0.85\textwidth]{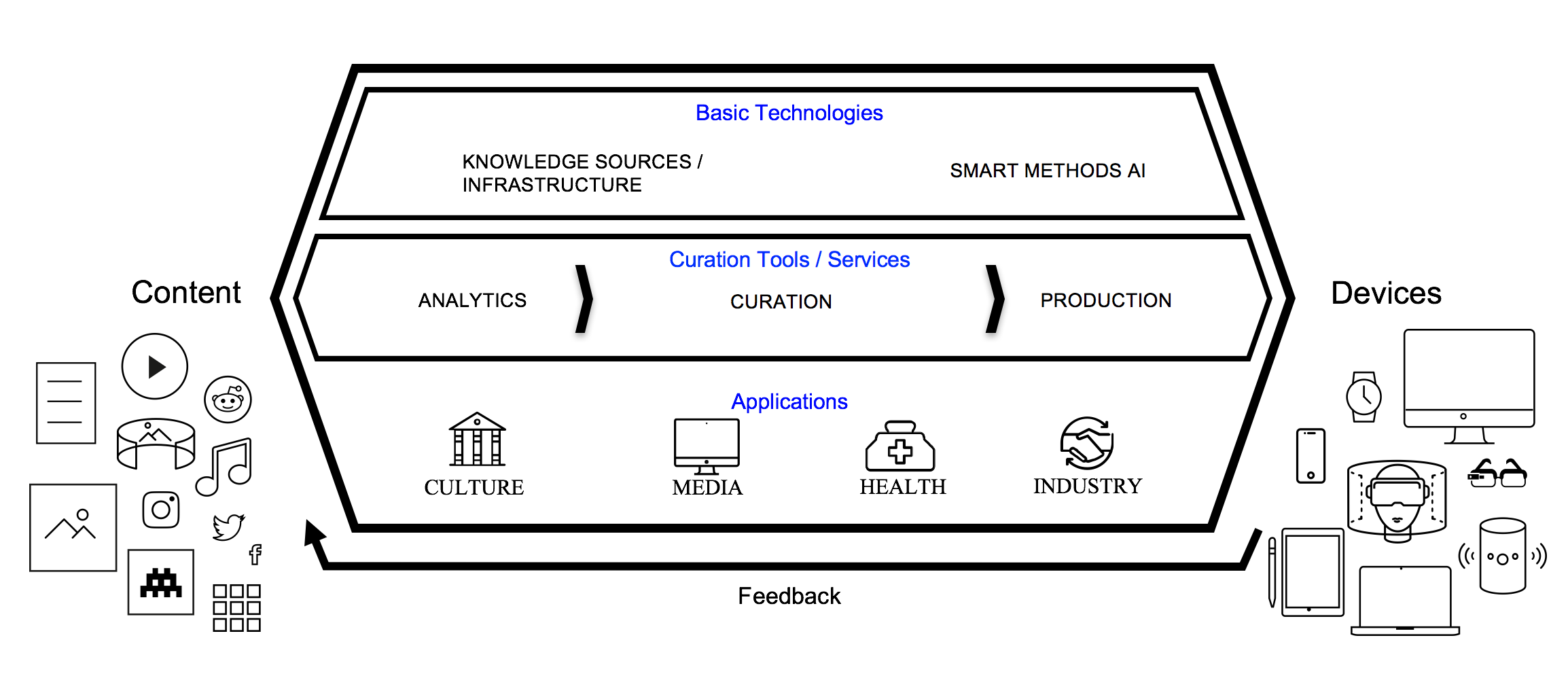}
\caption{Architectural concept of the QURATOR Curation Technology Platform} \label{fig_plattform}
\end{figure}

Figure~\ref{fig_plattform} outlines the concept of the QURATOR Curation Technology Platform which can be divided into three main layers. In order to process and transform incoming data, text and multimedia streams from different sources to device-adapted, publishable content, various groups of components, services and technologies are applied. First, the set of basic technologies includes adapters to data, content and knowledge sources, as well as infrastructural tools and smart AI methods for the acquisition, analysis and generation of content. Second, knowledge workers can make use of curation tools and services which have knowledge sources and intelligent procedures already integrated in order to process content. Third, there are selected use case and application areas (culture, media, health, industry), i.\,e., the respective integration of curation tools and services. Each of the three layers has already been populated with technology components which are to be further developed and also extended in the following two years of the project.

\section{QURATOR -- Partner Projects}
\label{sec:projects}

The project consortium includes ten partners: the research centers DFKI and Fraunhofer FOKUS, the industry partners 3pc, Ada Health, ART+COM, Condat, Semtation, Ubermetrics as well as Wikimedia Germany and the Berlin State Library (Stiftung Preu{\ss}ischer Kulturbesitz). Several of these partners already contributed to previous BMBF-funded projects including Corporate Smart Content\footnote{\url{https://www.unternehmen-region.de/de/7923.php}} and Digital Curation Technologies\footnote{\url{http://digitale-kuratierung.de}}, which focused on semiautomatic methods for the efficient processing, creation and distribution of high quality media content and laid the groundwork for the QURATOR project. 

In the following, we briefly introduce each partner and provide an overview of their respective projects focusing upon the current state and the next steps.

\subsection{DFKI GmbH: A Flexible AI Platform for
the Adaptive Analysis and Creative Generation of Digital Content}

DFKI (Deutsches Forschungszentrum f\"ur K\"unstliche Intelligenz GmbH) is Germany's leading research center in the field of innovative software technologies based on AI methods. The Speech and Language Technology Lab conducts advanced research in language technology and provides novel computational techniques for processing text, speech and knowledge.

In QURATOR, DFKI focuses on the development of an innovative platform for digital curation technologies \cite{rehm2016j,rehm2019b,rehm2018g,rehm2018f} as well as on the population of this platform with various processing services. This platform plays a crucial role in the project as it is being designed together with all partners who also contribute services to the platform.\footnote{This platform is developed in close collaboration with the EU project European Language Grid, which is also coordinated by DFKI, see \url{https://www.european-language-grid.eu} and \cite{elg2020} for more details.} Ultimately, the QURATOR platform will contain services, data sets and components that are able to handle and to process different types and classes of content as well as content sources. The DFKI services can be divided into three classes.

\emph{Preprocessing} encompasses the services that are responsible for obtaining and processing information from different content sources so that they can be used in the platform and integrated into other services \cite{rehm2018k}. These services include the provisioning of data sets and content (web pages, RSS feeds etc.), language and duplicate detection as well as document structure recognition.

\emph{Semantic analysis} includes services that process a document (or part of it) and add information in the form of annotations. These services are named entity recognition and linking, temporal expression analysis, relation extraction, event detection, fake news as well as discourse analysis \cite{rehm2016p,rehm2016q,rehm2017h,rehm2019e}.

\emph{Content generation} contains services that make use of annotated information (semantic analysis) to help create a new piece of information. These services are summarization, paraphrasing, automatic translation and semantic storytelling for both text and multimedia content \cite{rehm2019a,rehm2017m,rehm2017o,rehm2017b,rehm2017c,rehm2016n}.

DFKI will continue the development of the different services as well as the infrastructure. Since a flexible organization needs to be guaranteed, DFKI is also responsible for the design and implementation of workflows. These will ultimately enable the joint use of (almost) all the services available in the platform.

\subsection{3pc GmbH: Curation Technologies for Interactive Storytelling} 

3pc creates solutions for the digital age, combining strategy, design, technology, and communication in a holistic and user-centered approach. As experts in the development of novel and unique digital products, 3pc identifies core challenges and key content within complex subject matters.

\begin{figure}[ht]
\floatbox[{\capbeside\thisfloatsetup{capbesideposition={right,top},capbesidewidth=0.25\textwidth}}]{figure}[\FBwidth]
{\caption{Prototype of an asset management tool for analyzing and enriching multimedia files. The screenshot depicts a video analysis, creating semi-automated transcriptions, sections, clips, and thumbnails, as well as entity recognition and sentiment analysis.}
\label{fig_3pc-1}}
{\includegraphics[width=0.6\textwidth]{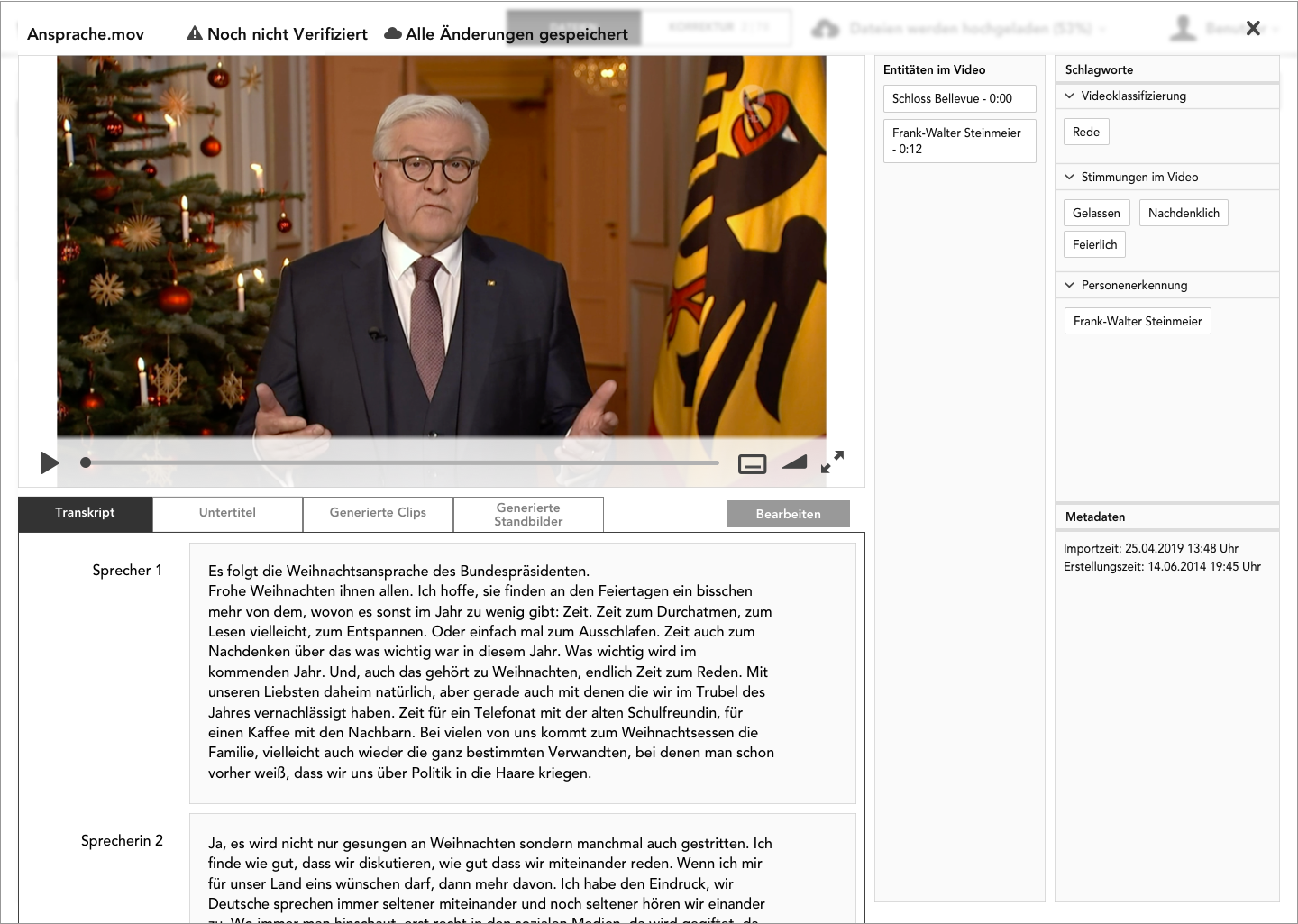}}
\end{figure}

As part of the QURATOR project, 3pc develops intelligent tools for interactive storytelling \cite{berger2017,rehm2017m} in order to assist editors, content curators, and publishers from cultural and scientific institutions, corporate communication divisions, and media agencies. The providers for storytelling face an increasing challenge telling engaging stories built from vast amounts of data for a broad range of devices, including wearables, augmented and virtual reality systems, voice-based user interfaces -- and whatever the future holds. In this context, interactive storytelling is defined as novel media formats and implementations that exceed today’s rather static websites by far. 3pc is currently building an asset management tool that enables users to access media analysis algorithms in an intuitive and efficient way (Figure~\ref{fig_3pc-1}). Media analysis processes text, images, videos, and audio files in order to enrich them with additional information such as content description, sentiment or topic, which is usually a labor-intensive and, therefore, expensive process often neglected in busy publishing environments. Enriched media becomes machine-readable, allowing storytellers to find content faster and for new connections to be forged in order to create richer, interactive stories.
Ultimately, a semantic storytelling machine becomes possible, generating semi-automatically unique and tailored stories, based entirely on user preferences. At 3pc, research is conducted through an iterative process by creating functional prototypes and testing their usefulness on representative members of different user groups. 3pc ensures that all novel technology solutions are adapted to each user’s needs, taking into consideration their tasks, behaviour and knowledge.

Next up, 3pc will extend traditional forms of interactive storytelling by exploring space, voice, and generated audio as means of human-computer interaction. Further research will also be conducted on training algorithms for domain-specific tasks in order to develop curation tools for different areas of expertise.

%\begin{figure}[ht]
%\floatbox[{\capbeside\thisfloatsetup{capbesideposition={right,top},capbesidewidth%=0.5\textwidth}}]{figure}[\FBwidth]
%{\caption{Prototype of an interactive storytelling website engaging users to %explore Kurt Tucholsky’s biography by offering several parallel storylines, %junctions and content alternatives to dive deeper into aspects of his life.}
%    \label{fig_3pc-2}}
%{\includegraphics[width=0.3\textwidth]{191003_3pc_storytelling-overview.png}}
%\end{figure}

\subsection{Ada Health GmbH: Curation of Biomedical Knowledge}

Ada Health GmbH is a global health company founded by doctors, scientists, and industry pioneers to create new possibilities for personal health. Ada's core system connects medical knowledge with intelligent technology to help people actively manage their health and medical professionals to deliver effective care.

Within the QURATOR project, Ada focuses on supporting the structured medical knowledge creation by providing a tool for pre-extracting information from unstructured text. This tool utilizes methods from biomedical Natural Language Processing (NLP). Since the quality of the medical database is of utmost importance to ensure accurate diagnosis support, the ``human in the loop'' approach leverages the deep medical knowledge provided by Ada's doctors and the efficiency of AI methods.

As a first step, Ada's researchers focus on the extraction of medical entities from medical case reports. These descriptions of a patient's symptoms are usually semi-structured and can function as test cases for Ada's quality control. The extraction of a structured case requires NLP methods such as the detection of relevant paragraphs, named entity recognition and named entity normalization. Challenging characteristics of named entities in the biomedical domain are their discontinuous nature in text as well as their high heterogeneity in terms of linguistic features. Thus, these domain-specific characteristics require the adaptation and implementation of domain-tailored NLP solutions. In order to do so, data is required which Ada acquires through a combination of manual annotation and active learning from feedback given by the medical content editors.

\subsection{ART+COM AG: Curation Tools for Multimedia Content}

ART+COM Studios designs and develops new media spaces and installations. New technology is not only used as an artistic medium of expression but as a medium for the interactive communication of complex information. In the process, ART+COM improves existing technologies constantly and explores their applications both independently and in cooperation with other companies and academic institutions.

The main focus within QURATOR is to develop basic technologies to automatically process and assess multimedia content and ultimately create smart exhibits that can organize and generate content automatically.

\begin{figure}[ht]
\floatbox[{\capbeside\thisfloatsetup{capbesideposition={right,top},capbesidewidth=0.45\textwidth}}]{figure}[\FBwidth]
{\caption{Prototype of the Wikidata knowledge graph tool displaying a selection of items and their connections to each other. The layout of the graphs is achieved using force-directed algorithms in combination with high-dimensional embeddings and manual curation.}
    \label{fig_ART+COM_1}}
{\includegraphics[width=0.5\textwidth]{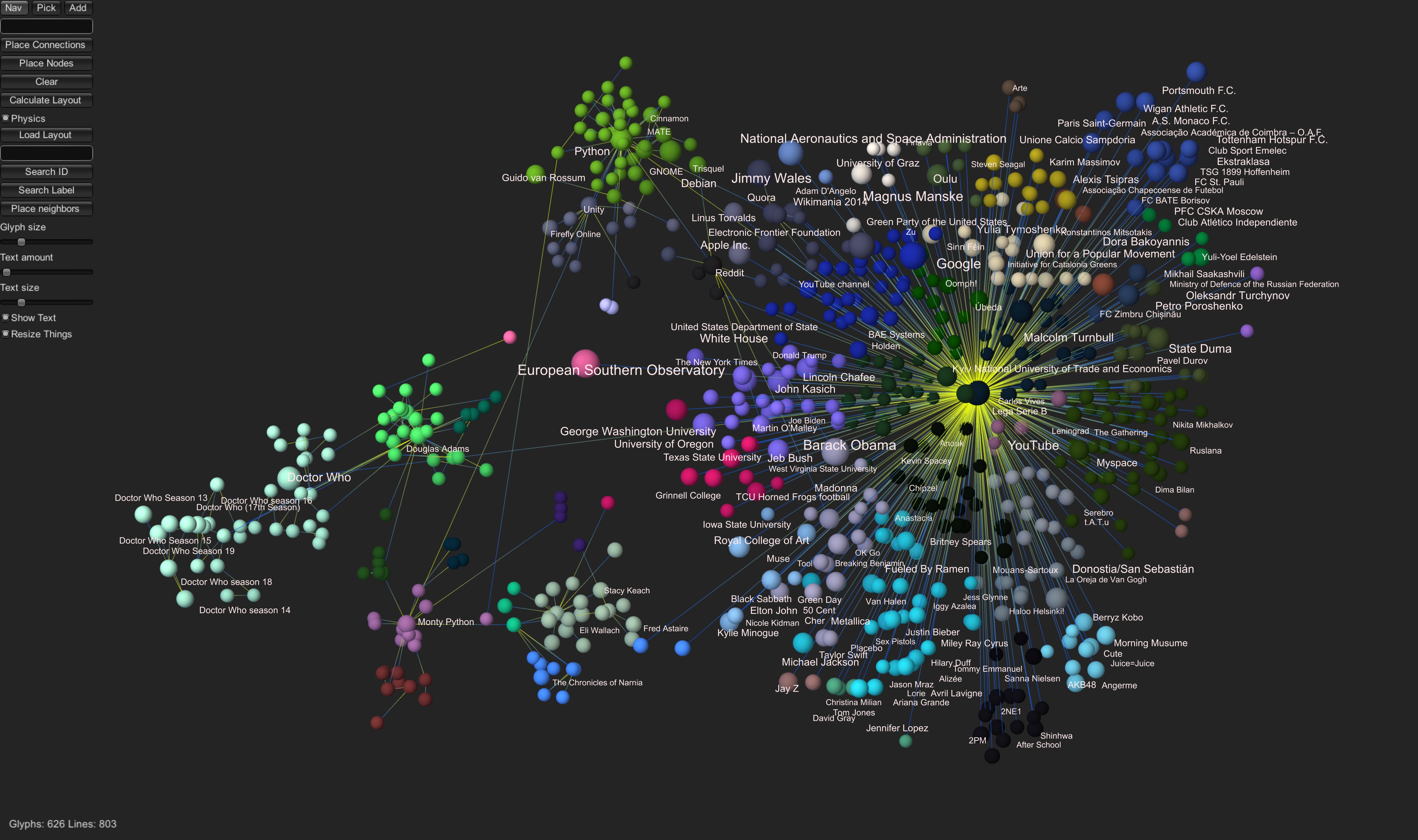}}
\end{figure}

The current objective is to categorize entities and visualize their relations in huge datasets. ART+COM's curation tool will import results from machine learning (ML) methods, including NLP, action detection, image recognition and processing of interconnected data. Subsequently, it should be used to create automated content analyses and visualizations that are navigable and support knowledge workers as well as the creation of interactive museum exhibits. The interconnected entities in Wikidata, one of the most relevant databases for researchers, are subject of one of the ongoing sub-projects. The project focuses on an interactive software to visualize, explore, and curate knowledge contained in Wikidata. Of particular interest are interaction techniques to filter and select the objects of interest and their relationships between each other. The software framework also explores potential data arrangements in two-dimensional and three-dimensional space, tailored to their relevance to particular questions and interests. Figure~\ref{fig_ART+COM_1} shows the prototype of the Wikidata knowledge graph tool displaying a selection of items and their connections to each other.

\subsection{Condat AG: Smart Newsboard}

Condat AG has a strong focus on the media industry, mainly on public broadcasters. Condat support all parts of the distribution chain for these broadcasters.

\begin{figure}
\includegraphics[width=0.7\textwidth]{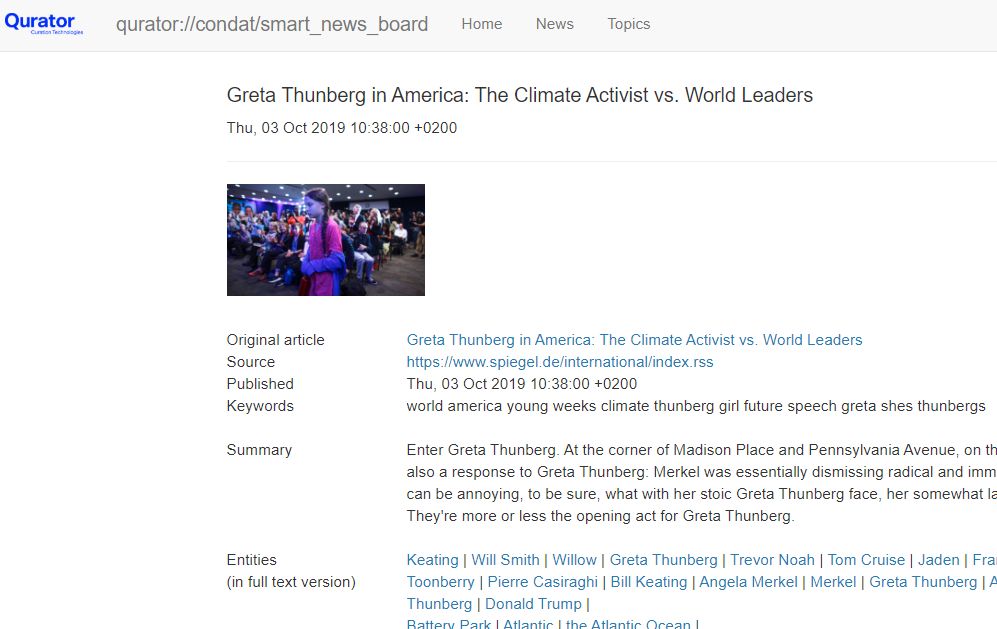}
\caption{Smart Newsboard to produce content based on news around a particular topic} \label{fig_condat-1}
\end{figure}

Within the QURATOR project, Condat develops a Smart Newsboard (Figure~\ref{fig_condat-1}) as a means to produce content based on news around a particular topic. This involves a chain of different curation services such as 1) finding the original sources around a topic by searches and subscription to RSS feeds, Twitter channels etc., 2) categorizing and classifying sources into meaningful groups, either through topic detection (if the categories are not predefined) or text classification (if the categories are fixed), or even a combination of both, 3) applying text analysis, named entity recognition and enrichment by linking the entities to specific resources such as Wikidata which allows the building of a knowledge graph to find connections between, e.\,g., people and events in different contexts and, thus, enable journalists to pursue a deeper analysis. Condat also explores the identification of temporal expressions which enables the possibility to generate story outlines based on the sequence of events as extracted from multiple documents (timelining). Another curation service relevant for the Smart Newsboard is the summarization of sources. This includes not only the summarization of individual texts but more importantly the summarization of multiple documents. 

Curation services for summarization, named entity recognition and topic detection have already been implemented. The next steps are the integration of additional services, as they become available, and the design of the user interface.

\subsection{Fraunhofer FOKUS: Corporate Smart Insights}

The Fraunhofer Institute for Open Communication Systems (FOKUS) develops solutions for the communication infrastructure of the future. With more than 30 years of experience, FOKUS is one of the most important actors in the ICT research landscape both nationally and worldwide. Fraunhofer FOKUS develops innovative processes from the original concept up to the pre-product in companies and institutions. As a member of important standardization bodies, the institute contributes to the definition of new ICT standards. It researches and develops application-orientated solutions for partners in industry, research and public administration in various ICT fields.

\begin{figure}
\includegraphics[width=0.7\textwidth]{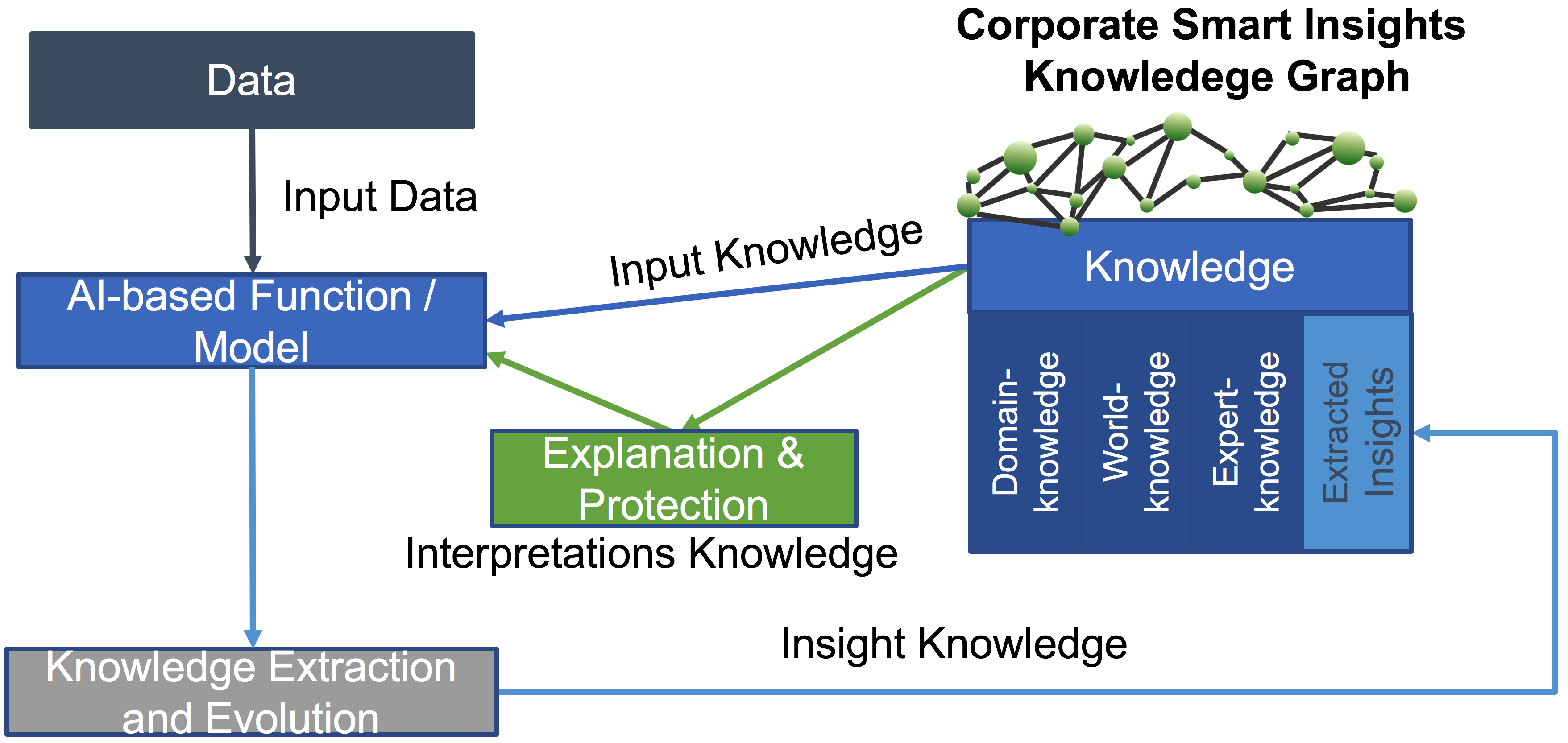}
\caption{The Corporate Smart Insights (CSI) concept.} \label{fig_fokus_csi_c}
\end{figure}

Fraunhofer FOKUS has significant experience in semantic data intelligence and AI, concentrated in the DANA group, which drives the research and the development in the area of Corporate Semantic Web. Using this experience, the aspect realized in the QURATOR project is an insight-driven AI approach (Figure~\ref{fig_fokus_csi_c}) which benefits the technological innovation of an Insight Driven Organisation (IDO).\footnote{\url{https://qurator.ai/partner/fraunhofer-fokus/}} An IDO embeds corporate knowledge, reasoning and smart insights learned from data analytics into the daily decisions and actions of an organisation, including their argumentation and interpretation. 

The technical CSI framework consists of knowledge repositories for the distributed management of knowledge artefacts, such as semantic knowledge graphs and terminologies, a standardized API for  knowledge bases  \cite{paschke2015representational}, a knowledge extraction and analytics\footnote{\url{https://www.cyber-akademie.de/anlage.jsp?id=959}} service, and methods for corporate smart insights knowledge evolution, as well as services to reuse the learned CSI knowledge for AI, including inference, explainability and plausibility/validation.

\subsection{Semtation GmbH: Intelligent Business Process Modelling}

Semtation GmbH provides the platform SemTalk (registered trademark) for modeling and supporting business processes and knowledge structures, an easy to use but very powerful modeling and portal tool based on Microsoft Visio and the Microsoft Cloud. SemTalk technology makes use of various tools provided in Microsoft 365 in order to offer best-in-class portal experiences when it comes to supporting business processes.

In QURATOR, Semtation pursues the enhancement of business process model usage. It consists of several tasks that aim in two directions, namely to 1) present models on other devices but a monitor and 2) recommend information dynamically based on the process context of the current user. Integrating various AI technologies is necessary to obtain suitable results for both scenarios. It helps to recognize your surroundings in order to recommend adequate information in mixed reality settings and to understand natural language in a chat scenario. It also makes it easier to understand the current process context in order to recommend available documents and team members in various settings based on text analysis and other machine learning use cases. Semtation has already integrated knowledge graph information in process portals in order to use available internal information to recommend suitable documents and people based on the current process or project instance and the current task.

The next step will be to define a scenario with one of the customers in order to check requirements and results on a real world basis.

\subsection{Stiftung Preu{\ss}ischer Kulturbesitz, Staatsbibliothek zu Berlin: Automated Curation Technologies for Digitized Cultural Heritage}

The Berlin State Library (Staatsbibliothek zu Berlin, SBB) is the largest research library in Germany with more than 12 million documents in its holdings and more than 2.5~PB of digital data stored throughout various repositories (as of Oct.~2019). The collection encompasses texts, media and cultural works from all fields in all languages, from all time periods and all countries of the world.

\begin{figure}
\centering
\includegraphics[width=0.38\textwidth]{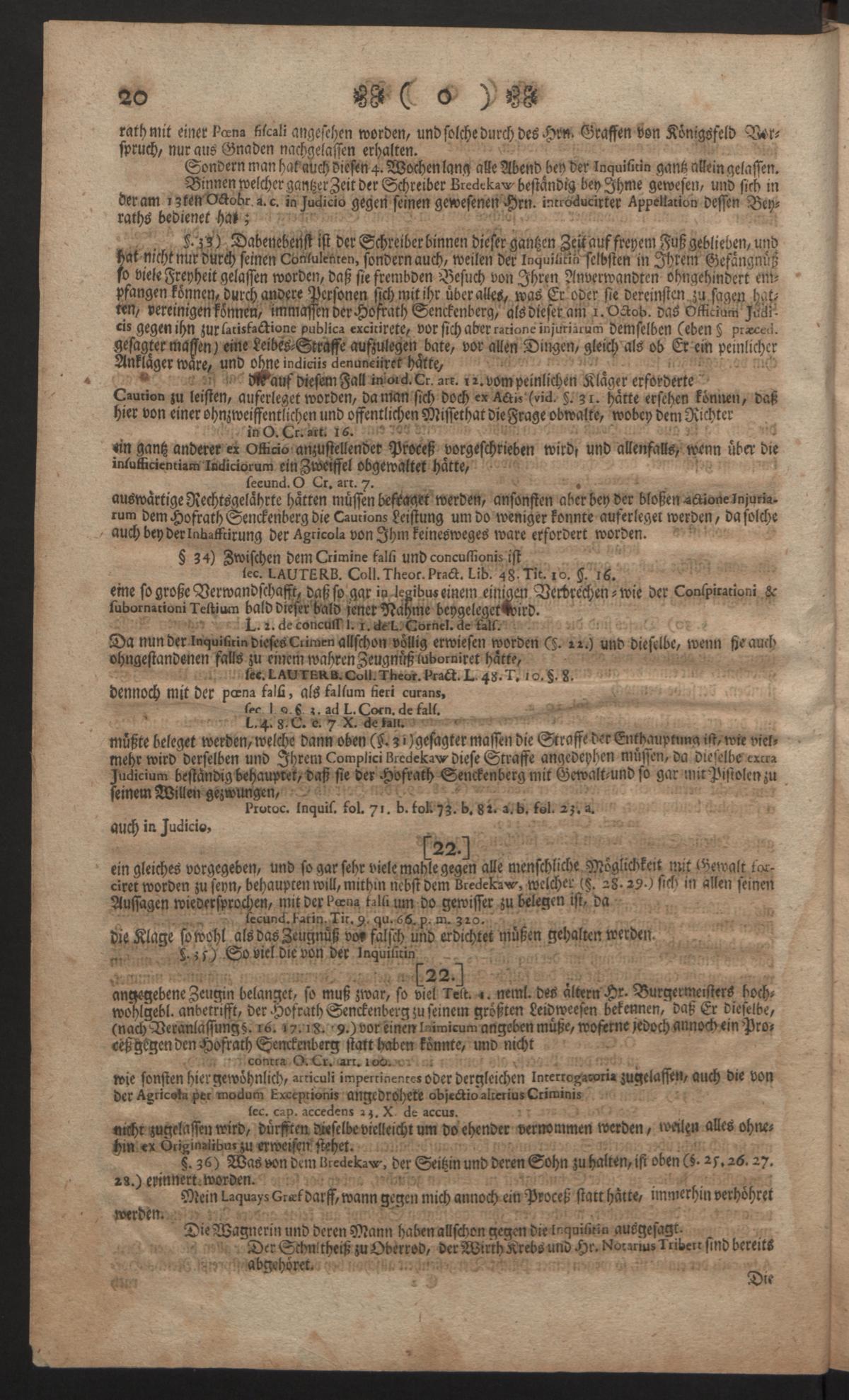}
\caption{Example image from a digitized collection} \label{SBB_digitized_collections}
\end{figure}

Within QURATOR, the Berlin State Library is taking part in the R\&D activities on behalf of the Prussian Heritage Foundation (Stiftung Preu{\ss}ischer Kulturbesitz, SPK). SBB aims to digitize all its copyright-free historical collections and to make them available on the web\footnote{\url{https://digital.staatsbibliothek-berlin.de}} as facsimile images with high-quality text, logically structured and semantically annotated for use by researchers. In order to achieve this goal, SBB works in a number of research areas in the context of QURATOR -- from layout and text recognition (OCR) and unsupervised post-correction to named entity recognition (NER), disambiguation and linking. Due to the huge volume and variety of the documents published between 1475 and 1945, solutions are required that are particularly robust and that can be fine-tuned to the complexities of historical fonts, layout, language and orthography. While the SBB adopts state-of-the-art convolutional neural networks (CNN) like ResNet50 \cite{he2015deep} and UNet \cite{ronneberger2015unet} in combination with attention and adds rule-based domain adaptation for layout recognition and the classification of structural elements, it follows a more classical RNN-LSTM-CTC approach for text recognition \cite{wick2018calamari}, achieving character error rates below 1\% with voting between multiple models trained on sufficiently large amounts of historical document ground truth data \cite{springmann2018ground}. In a related effort, SBB is also taking part in the development of an open end-to-end framework for historical document analysis and recognition based on AI methods \cite{neudecker2019}. For NER, the recent transformer architecture BERT \cite{devlin2018bert} is utilized and adapted to historical German through a combination of unsupervised pre-training and supervised learning \cite{labusch2019}. The final goal is to identify and classify named entities found in the digitized documents, and to disambiguate and link them to an online knowledge base, e.\,g., Wikidata.

Eventually, the digitized historical documents shall be made fully searchable with semantic markup enabling advanced content retrieval scenarios and rich contextualization of documents with knowledge from third party sources. In a further step, image-based classification methods will be added to enhance the document metadata and to complement the functionalities offered through the full-text search. In the course of 2020, a demonstrator will be launched in SBB's research and innovation lab.\footnote{\url{https://lab.sbb.berlin}}

\subsection{Ubermetrics Technologies GmbH: Curation Technologies for the Monitoring of Online Content and Risks}

\begin{figure}
\includegraphics[width=0.50\textwidth]{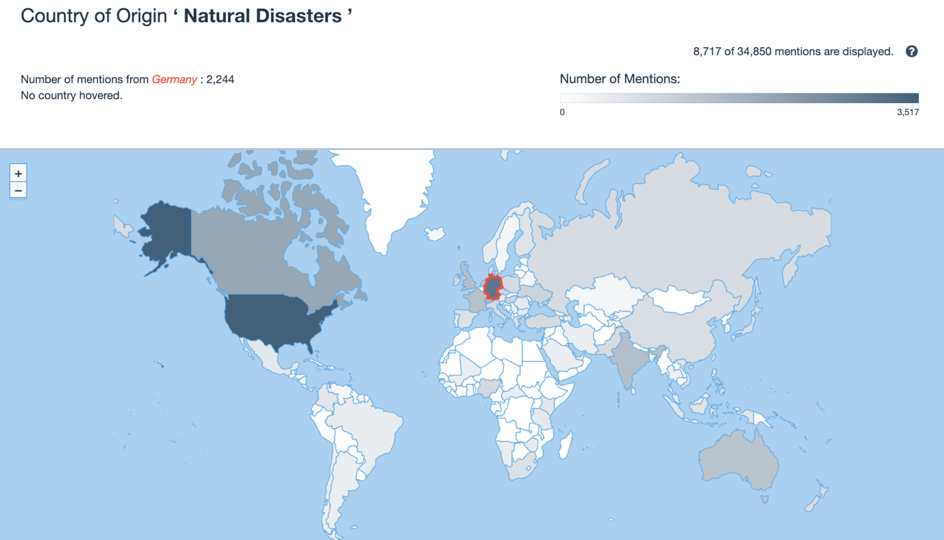}
\caption{Analysis of mentions of natural disasters worldwide} \label{fig_ubermetrics-1}
\end{figure}

Ubermetrics is a leading provider of cloud-based social media monitoring and content intelligence software. Ubermetrics analyses public data from online, print, TV and radio sources with a proprietary technology to identify critical information in order to help organizations to optimize decision processes and increase their performance (see Figure~\ref{fig_ubermetrics-1}).

In QURATOR, Ubermetrics researches how to use social media for the monitoring of both external and internal risks. The focus areas are an easy setup of risk-related search queries thanks to automated query suggestions and a condensation of the results found with the help of summarization and duplicate detection technology. The project aims at showing the developed capabilities in demonstrators to get feedback for a later product for risk monitoring.

Automatic connections to risk related sources have already been developed and a first version of query suggestions is available. The next steps are to improve the query suggestions especially in the risk context and start with the research and development of text summarization methods.

%\begin{figure}
%\includegraphics[width=\textwidth]{Ubermetrics_query_for_finding_mentions_of_problematic_working_conditions.png}
%\caption{Ubermetrics – query for finding mentions of problematic working conditions} %\label{fig_ubermetrics-2}
%\end{figure}

\subsection{Wikimedia Deutschland e.\,V.: Data quality in Wikidata}

Wikidata is Wikimedia's knowledge base. It is a sister project of Wikipedia and collects general purpose data about the world. Wikidata currently describes over 63 million entities such as people, geographic entities, events and works of art. Wikidata, just like Wikipedia, is built by a community --~currently consisting of more than 20,000 editors from all around the world~-- that collects and maintains that data. Wikidata's data powers a large number of applications, among them search engine instant answers, digital personal assistants, educational websites, as well as information boxes on Wikipedia articles. By its nature, Wikidata is an open project. It relies on contributions from volunteer editors. To build a large enough community for building and maintaining a general purpose knowledge base the entry barrier needs to be low. At the same time the pressure to provide high-quality data is increasing as more and more people are exposed to its data in their day-to-day life. It is vital for the long-term sustainability of Wikidata to find ways to stay open while keeping the quality of its data high. On top of that Wikidata can only follow its mission of giving more people more access to more knowledge if the data is easily accessible for everyone. In QURATOR, we work on improving both the quality and accessibility of the data in Wikidata, which is supposed to become a viable basic building block of the QURATOR platform by providing easily accessible high-quality data for all partners.

\begin{figure}
\includegraphics[width=0.5\textwidth]{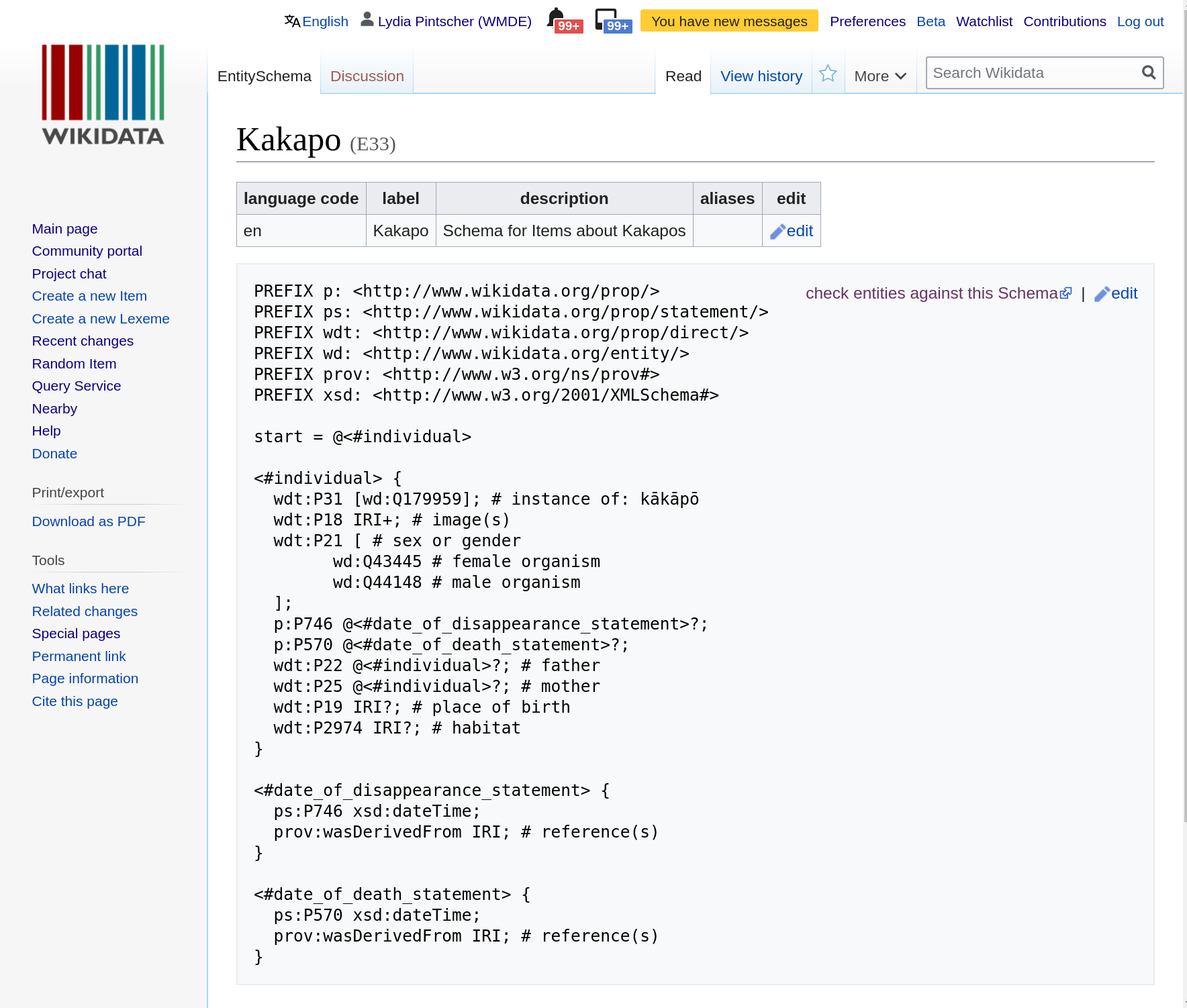}
\caption{Schema for describing Kakapos in Wikidata}
\label{fig_Wikidata-1}
\end{figure}

So far three important components with a focus on quality improvements have been developed. The first presents a way to define schemas for data in order to allow editors to quickly find data that does not conform to the specified schema. It is based on the Shape Expression standard (see Figure~\ref{fig_Wikidata-1}). The second entails the ability to automatically judge the quality of a data item using machine learning. Editors can then find especially high and low quality data items to showcase and improve them respectively. The third is an improved way to add references for individual data points to improve the verifiability of the data.

\section{Summary and Next Steps}
\label{sec:summary}

This paper provides a snapshot of the technologies and approaches developed as part of the QURATOR project. Its vision is to offer a broad portfolio of integrated solutions to the cross-industry challenges that are associated with the curation of digital content. A platform strategy has been developed to transform fragmented market areas for curation technologies into a new stand-alone market and greatly expand it by displacing existing isolated solutions. QURATOR aims to establish an ecosystem for curation technologies that improve the state of the art and transform the Berlin-Brandenburg area into a global center of excellence for curation technologies and the development of efficient industrial applications.

\section*{Acknowledgements}
The research presented in this article is funded by the German Federal Ministry of Education and Research (BMBF) through the project QURATOR (Unternehmen Region, Wachstumskern, grant no.~03WKDA1A). \url{http://qurator.ai}

\bibliographystyle{./splncs04}
\bibliography{./bibliography}

\begin{thebibliography}{10}
\providecommand{\url}[1]{\texttt{#1}}
\providecommand{\urlprefix}{URL }
\providecommand{\doi}[1]{https://doi.org/#1}

\bibitem{berger2017}
Berger, A.: {Archive zum Sprechen Bringen -- Semantic Storytelling oder der
  Redaktionsworkflow der Zukunft}. In: {23. Berliner Veranstaltung der
  Internationalen EVA-Serie Electronic Media and Visual Arts}. pp. 135--141
  (2017)

\bibitem{rehm2016j}
Bourgonje, P., Moreno-Schneider, J., Nehring, J., Rehm, G., Sasaki, F.,
  Srivastava, A.: {Towards a Platform for Curation Technologies: Enriching Text
  Collections with a Semantic-Web Layer}. In: Sack, H., Rizzo, G., Steinmetz,
  N., MladeniÄ‡, D., Auer, S., Lange, C. (eds.) The Semantic Web. pp.
  65--68. No.~9989 in Lecture Notes in Computer Science, Springer (June 2016),
  eSWC 2016 Satellite Events. Heraklion, Crete, Greece, May 29 -- June 2, 2016
  Revised Selected Papers

\bibitem{rehm2016p}
Bourgonje, P., Schneider, J.M., Rehm, G., Sasaki, F.: {Processing Document
  Collections to Automatically Extract Linked Data: Semantic Storytelling
  Technologies for Smart Curation Workflows}. In: Gangemi, A., Gardent, C.
  (eds.) Proceedings of the 2nd International Workshop on Natural Language
  Generation and the Semantic Web (WebNLG~2016). pp. 13--16. The Association
  for Computational Linguistics, Edinburgh, UK (September 2016)

\bibitem{devlin2018bert}
Devlin, J., Chang, M.W., Lee, K., Toutanova, K.: {Bert: Pre-training of Deep
  Bidirectional Transformers for Language Understanding}. arXiv preprint
  arXiv:1810.04805  (2018)

\bibitem{he2015deep}
He, K., Zhang, X., Ren, S., Sun, J.: {Deep Residual Learning for Image
  Recognition}. In: Proceedings of the IEEE Conference on Computer Vision and
  Pattern Recognition. pp. 770--778 (2016)

\bibitem{labusch2019}
Labusch, K., Neudecker, C., Zellh\"{o}fer, D.: {BERT for Named Entity
  Recognition in Contemporary and Historic German}. In: Preliminary Proceedings
  of the 15th Conference on Natural Language Processing (KONVENS 2019): Long
  Papers. pp.~1--9. German Society for Computational Linguistics \& Language
  Technology, Erlangen, Germany (2019)

\bibitem{rehm2017o}
Moreno-Schneider, J., Srivastava, A., Bourgonje, P., Wabnitz, D., Rehm, G.:
  {Semantic Storytelling, Cross-lingual Event Detection and other Semantic
  Services for a Newsroom Content Curation Dashboard}. In: Popescu, O.,
  Strapparava, C. (eds.) Proc. of the Second Workshop on Natural Language
  Processing meets Journalism -- EMNLP 2017 Workshop (NLPMJ~2017). pp. 68--73.
  Copenhagen, Denmark (2017)

\bibitem{neudecker2019}
Neudecker, C., Baierer, K., Federbusch, M., Boenig, M., W\"{u}rzner, K.M.,
  Hartmann, V., Herrmann, E.: {OCR-D: An End-to-end Open Source OCR Framework
  for Historical Printed Documents}. In: Proceedings of the 3rd International
  Conference on Digital Access to Textual Cultural Heritage. pp. 53--58.
  DATeCH2019, ACM, New York, NY, USA (2019). \doi{10.1145/3322905.3322917},
  \url{http://doi.acm.org/10.1145/3322905.3322917}

\bibitem{rehm2019e}
Ostendorff, M., Bourgonje, P., Berger, M., Moreno-Schneider, J., Rehm, G.:
  {Enriching BERT with Knowledge Graph Embeddings for Document Classification}.
  In: Remus, S., Aly, R., Biemann, C. (eds.) Proceedings of the GermEval
  Workshop 2019 -- Shared Task on the Hierarchical Classification of Blurbs.
  Erlangen, Germany (10 2019), 8 October 2019

\bibitem{paschke2015representational}
Paschke, A., Athan, T., Sottara, D., Kendall, E., Bell, R.: {A Representational
  Analysis of the API4KP Metamodel}. In: International Workshop Formal
  Ontologies Meet Industries. pp. 1--12. Springer (2015)

\bibitem{elg2020}
Rehm, G., Berger, M., Elsholz, E., Hegele, S., Kintzel, F., Marheinecke, K.,
  Piperidis, S., Deligiannis, M., Galanis, D., Gkirtzou, K., Labropoulou, P.,
  Bontcheva, K., Jones, D., Roberts, I., Hajic, J., Hamrlova, J., Kacena, L.,
  Choukri, K., Arranz, V., Mapelli, V., Vasiljevs, A., Anvari, O., Lagzdins,
  A., Melnika, J., Backfried, G., Dikici, E., Janosik, M., Prinz, K., Prinz,
  C., Stampler, S., Thomas-Aniola, D., Perez, J.M.G., Silva, A.G., Berrio, C.,
  Germann, U., Renals, S., Klejch, O.: {European Language Grid: An Overview}
  (2020), submitted to LREC 2020. Marseille, France.

\bibitem{rehm2017b}
Rehm, G., He, J., Schneider, J.M., Nehring, J., Quantz, J.: {Designing User
  Interfaces for Curation Technologies}. In: Yamamoto, S. (ed.) Human Interface
  and the Management of Information: Information, Knowledge and Interaction
  Design, 19th International Conference, HCI International 2017 (Vancouver,
  Canada). pp. 388--406. No. 10273 in Lecture Notes in Computer Science (LNCS),
  Springer, Cham, Switzerland (July 2017), part I

\bibitem{rehm2019b}
Rehm, G., Lee, M., Schneider, J.M., Bourgonje, P.: {Curation Technologies for a
  Cultural Heritage Archive: Analysing and Transforming a Heterogeneous Data
  Set into an Interactive Curation Workbench}. In: Antonacopoulos, A., Bechler,
  M. (eds.) Proceedings of DATeCH 2019: Digital Access to Textual Cultural
  Heritage. Brussels, Belgium (May 2019), 8-10 May 2019. In print.

\bibitem{rehm2015c}
Rehm, G., Sasaki, F.: {Digitale Kuratierungstechnologien -- Verfahren f\"ur die
  Effiziente Verarbeitung, Erstellung und Verteilung Qualitativ Hochwertiger
  Medieninhalte}. In: Proceedings der Fr\"uhjahrstagung der Gesellschaft f\"ur
  Sprachtechnologie und Computerlinguistik (GSCL~2015). pp. 138--139. Duisburg
  (9 2015), 30.~September--2.~Oktober

\bibitem{rehm2017m}
Rehm, G., Schneider, J.M., Bourgonje, P., Srivastava, A., Fricke, R., Thomsen,
  J., He, J., Quantz, J., Berger, A., K\"{o}nig, L., R\"{a}uchle, S., Gerth,
  J., Wabnitz, D.: {Different Types of Automated and Semi-Automated Semantic
  Storytelling: Curation Technologies for Different Sectors}. In: Rehm, G.,
  Declerck, T. (eds.) Language Technologies for the Challenges of the Digital
  Age: 27th International Conference, GSCL 2017, Berlin, Germany, September
  13-14, 2017, Proceedings. pp. 232--247. No. 10713 in Lecture Notes in
  Artificial Intelligence (LNAI), Gesellschaft f\"{u}r Sprachtechnologie und
  Computerlinguistik e.V., Springer, Cham, Switzerland (January 2018),
  13/14~September 2017.

\bibitem{rehm2017h}
Rehm, G., Schneider, J.M., Bourgonje, P., Srivastava, A., Nehring, J., Berger,
  A., K\"{o}nig, L., R\"{a}uchle, S., Gerth, J.: {Event Detection and Semantic
  Storytelling: Generating a Travelogue from a large Collection of Personal
  Letters}. In: Caselli, T., Miller, B., van Erp, M., Vossen, P., Palmer, M.,
  Hovy, E., Mitamura, T. (eds.) Proc. of the Events and Stories in the News
  Workshop. pp. 42--51. Association for Computational Linguistics, Vancouver,
  Canada (August 2017)

\bibitem{rehm2019a}
Rehm, G., Zaczynska, K., Schneider, J.M.: {Semantic Storytelling: Towards
  Identifying Storylines in Large Amounts of Text Content}. In: Jorge, A.,
  Campos, R., Jatowt, A., Bhatia, S. (eds.) Proc. of Text2Story -- Second
  Workshop on Narrative Extraction From Texts co-located with 41th European
  Conf. on Information Retrieval (ECIR 2019). pp. 63--70. Cologne, Germany
  (April 2019), 14 April 2019

\bibitem{ronneberger2015unet}
Ronneberger, O., Fischer, P., Brox, T.: {U-net: Convolutional Networks for
  Biomedical Image Segmentation}. In: {International Conference on Medical
  Image Computing and Computer-assisted Intervention}. pp. 234--241. Springer
  (2015)

\bibitem{rehm2016n}
Schneider, J.M., Bourgonje, P., Nehring, J., Rehm, G., Sasaki, F., Srivastava,
  A.: {Towards Semantic Story Telling with Digital Curation Technologies}. In:
  Birnbaum, L., Popescu, O., Strapparava, C. (eds.) Proceedings of Natural
  Language Processing Meets Journalism -- IJCAI-16 Workshop (NLPMJ~2016). New
  York (July 2016)

\bibitem{rehm2017c}
Schneider, J.M., Bourgonje, P., Rehm, G.: {Towards User Interfaces for Semantic
  Storytelling}. In: Yamamoto, S. (ed.) Human Interface and the Management of
  Information: Information, Knowledge and Interaction Design, 19th Int. Conf.,
  HCI International 2017 (Vancouver, Canada). pp. 403--421. No. 10274 in
  Lecture Notes in Computer Science (LNCS), Springer, Cham, Switzerland (July
  2017), part II

\bibitem{rehm2018g}
Schneider, J.M., Rehm, G.: {Curation Technologies for the Construction and
  Utilisation of Legal Knowledge Graphs}. In: Rehm, G., Rodriguez-Doncel, V.,
  Schneider, J.M. (eds.) Proc. of the LREC 2018 Workshop on Language Resources
  and Technologies for the Legal Knowledge Graph. pp. 23--29. Miyazaki, Japan
  (May 2018)

\bibitem{rehm2018f}
Schneider, J.M., Rehm, G.: {Towards a Workflow Manager for Curation
  Technologies in the Legal Domain}. In: Rehm, G., Rodriguez-Doncel, V.,
  Schneider, J.M. (eds.) Proc. of the LREC 2018 Workshop on Language Resources
  and Technologies for the Legal Knowledge Graph. pp. 30--35. Miyazaki, Japan
  (May 2018)

\bibitem{rehm2018k}
Schneider, J.M., Roller, R., Bourgonje, P., Hegele, S., Rehm, G.: {Towards the
  Automatic Classification of Offensive Language and Related Phenomena in
  German Tweets}. In: Ruppenhofer, J., Siegel, M., Wiegand, M. (eds.)
  Proceedings of the GermEval Workshop 2018 -- Shared Task on the
  Identification of Offensive Language. pp. 95--103. Vienna, Austria (September
  2018), 21 September 2018

\bibitem{springmann2018ground}
Springmann, U., Reul, C., Dipper, S., Baiter, J.: {Ground Truth for Training
  OCR Engines on Historical Documents in German Fraktur and Early Modern
  Latin}. arXiv preprint arXiv:1809.05501  (2018)

\bibitem{rehm2016q}
Srivastava, A., Sasaki, F., Bourgonje, P., Moreno-Schneider, J., Nehring, J.,
  Rehm, G.: {How to Configure Statistical Machine Translation with Linked Open
  Data Resources}. In: Esteves-Ferreira, J., Macan, J., Mitkov, R., Stefanov,
  O.M. (eds.) Proceedings of Translating and the Computer 38 (TC38). pp.
  138--148. Editions Tradulex, London, UK (November 2016),
  \url{http://www.asling.org/tc38/}

\bibitem{wick2018calamari}
Wick, C., Reul, C., Puppe, F.: {Calamari-A High-Performance Tensorflow-based
  Deep Learning Package for Optical Character Recognition}. arXiv preprint
  arXiv:1807.02004  (2018)

\end{thebibliography}

\end{document}